\documentstyle[aps%,twocolumn%
,epsfig%
,preprint%
]{revtex}
\begin{document}
\draft
\preprint{Guchi-TP-001}
\date{\today%April 1998
}
\title{
Black Holes with Scalar Hair in $(2+1)$ dimensions
}
\author{Yoshitaka~Degura${}^1$,
 Kenji~Sakamoto${}^1$,
Kiyoshi~Shiraishi${}^{1,2}$%
\thanks{e-mail: {\tt g00345@simail.ne.jp,
shiraish@sci.yamaguchi-u.ac.jp}
% \hfill\break
}}
\address{${}^1$Graduate School of Science and Engineering, 
Yamaguchi University\\ 
Yoshida, Yamaguchi-shi, Yamaguchi 753-8512, Japan
}
\address{${}^2$Faculty of Science, Yamaguchi University\\
Yoshida, Yamaguchi-shi, Yamaguchi 753-8512, Japan
}
\maketitle
\begin{abstract}
%%%%%%%%%%%%%%%%%%%%%%%%%%%%%%%%%%%%%%%%%%%%%%%%%%%%%%%%%%%%%%%%%%%%%%
Nonrotating and rotating black hole soltuions 
in $(2+1)$ dimensions are studied in a model including
a real scalar field with a simple potential coupled to gravity.
%%%%%%%%%%%%%%%%%%%%%%%%%%%%%%%%%%%%%%%%%%%%%%%%%%%%%%%%%%%%%%%%%%%%%%
\end{abstract}

%\vspace{7mm}
%\vfill
%\eject

%%%%%%%%%%%%%%%%%%%%%%%%%%%%%%%%%%%%%%%%%%%%%%%%%%%%%%%%%%%%%%%%%%%%%%
%%%%%%%%%%%%%%%%%%%%%%%%%%%%%%%%%%%%%%%%%%%%%%%%%%%%%%%%%%%%%%%%%%%%%%
\section{Introduction}
%%%%%%%%%%%%%%%%%%%%%%%%%%%%%%%%%%%%%%%%%%%%%%%%%%%%%%%%%%%%%%%%%%%%%%
%%%%%%%%%%%%%%%%%%%%%%%%%%%%%%%%%%%%%%%%%%%%%%%%%%%%%%%%%%%%%%%%%%%%%%

In recent years, gravity in three dimensions has attracted much 
attention. Since Ba\~{n}ados, Teitelboim and Zanelli (BTZ) found
circularly-symmetric black hole solutions for three dimensional 
gravity with a negative cosmological constant~\cite{BTZ}, properties 
of their black holes~\cite{Car} and many other types of 
three dimensional black holes have been investigated.

On the other hand, static black hole solutions including matter fields 
in $(3+1)$ dimensions have been examined by many authors.
One of their examples is the Yang-Mills black hole in four dimensional 
spacetime~\cite{Biz}. 
Its properties have been studied and some descendants
have been considered recently~\cite{Tor}.

Static black holes with a nontrivial scalar field as a source of 
gravity is, however, problematic in four dimensions.
Beckenstein established a no-go theorem for static spherical black 
holes with such ``scalar hair'' in $(3+1)$ dimensions~\cite{Bek}.

Again we turn to the three dimensional case.
Since BTZ used a negative cosmological constant
and found the asymptotically no-flat black hole solution,
we do not have to restrict ourselves to the positive definite scalar 
potentials.
Thus the no-go theorem of Bekenstein will be evaded in the three 
dimensional case.
Actually, three-dimensional black hole slutions with scalar fields are
studied in various contexts~\cite{Mann,MZ,Chan}. 

%%%%%%%%%%%%%%%%%%%%%%%%%%%%%%%%%%%%%%%%%%%%%%%%%%%%%%%%%%%%%%%%%%%%%%

In the present paper,
we considered a simple class of scalar potentials in three dimensional 
gravity and construct circularly-symmetric black hole solutions in the
model.
In three dimensions, rotating solutions are most easily treated.
Therefore we also consider circularly symmetric rotating black holes 
with the scalar field.

%%%%%%%%%%%%%%%%%%%%%%%%%%%%%%%%%%%%%%%%%%%%%%%%%%%%%%%%%%%%%%%%%%%%%%

The action of our model is
\begin{equation}
S=\int d^{3}x\sqrt{-g}\left[\frac{1}{16\pi G}R -
\frac{1}{2} (\nabla \phi)^{2} - V(\phi) \right],
\label{eq:action}
\end{equation}
where $R$ is the scalar curvature and $\phi$ stands for a real scalar 
field. $G$ is the Newton's gravitational constant.

\begin{figure}[h]
\centering
\epsfbox{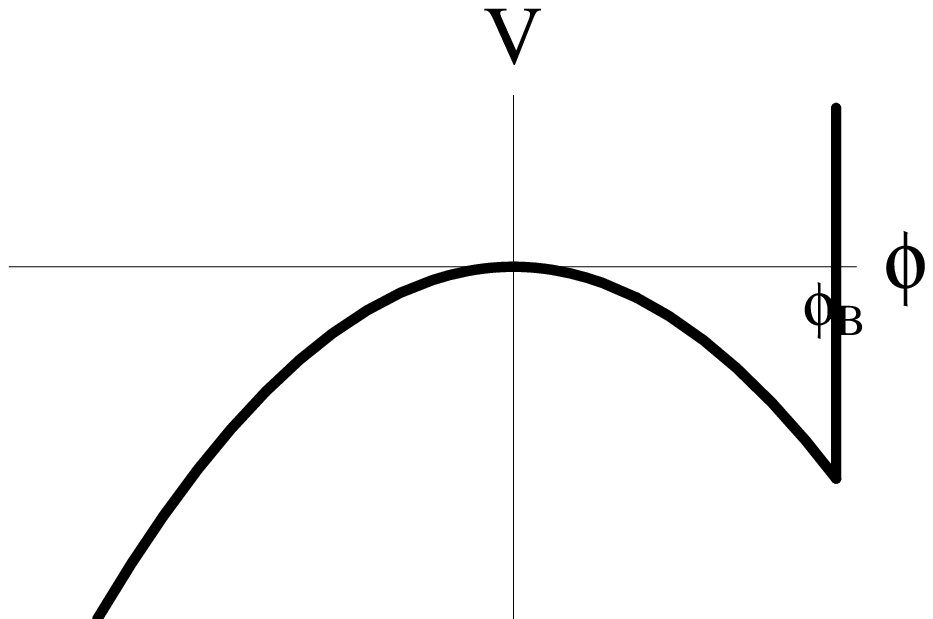}
\caption{A schematic view of the scalar potential.}
\label{fig:1}
\end{figure}

We assume that the potential $V(\phi)$ takes the form:
\begin{equation}
V(\phi)=\left\{\begin{array}{ccc}
-\frac{1}{2}\mu^2\phi^2 &~~~& \phi<\phi_B ,\\
\infty &~~~& \phi>\phi_B .\end{array}\right.
\end{equation}
A schematic view of the potential is given in FIG.~\ref{fig:1}.

In the far region from the black hole, $\phi$ falls into the bottom 
of the potential,
$\phi=\phi_B$. Therefore in the asymptotic region, the spacetime is 
the BTZ black hole spacetime with an effective negative cosmological
constant $-\lambda=-(8\pi G)\mu^2\phi_B^2/2$.

In Sec.~\ref{sec:2} in the present paper,
we derive nonrotating circularly-symmetric solutions in the model.
Rotating black holes are analyzed in this model in Sec.~\ref{sec:3}.
The last section~\ref{sec:4} is devoted to a brief conclusion.

%%%%%%%%%%%%%%%%%%%%%%%%%%%%%%%%%%%%%%%%%%%%%%%%%%%%%%%%%%%%%%%%%%%%%%
%%%%%%%%%%%%%%%%%%%%%%%%%%%%%%%%%%%%%%%%%%%%%%%%%%%%%%%%%%%%%%%%%%%%%%
\section{Nonrotating black holes with scalar hair}
\label{sec:2}
%%%%%%%%%%%%%%%%%%%%%%%%%%%%%%%%%%%%%%%%%%%%%%%%%%%%%%%%%%%%%%%%%%%%%%
%%%%%%%%%%%%%%%%%%%%%%%%%%%%%%%%%%%%%%%%%%%%%%%%%%%%%%%%%%%%%%%%%%%%%%

For the static, nonrotating, circularly-symmetric solutions,
the metric can be taken as 
\begin{equation}
ds^{2}= - e^{-2\delta(r)}\Delta(r)dt^{2} + \frac{1}{\Delta(r)} dr^{2}+
r^2d\theta^2.
\label{eq:met0}
\end{equation}

We also assume $\phi$ is a function only of the radial coodinate $r$.

One can obtain the field equations by varying
the action~(\ref{eq:action}).
When the assumption above is taken into consideration,
the field equations can be written as
\begin{eqnarray}
\frac{1}{2r}\frac{d\Delta}{dr}&=&8\pi G\left[-\frac{1}{2}\Delta\left(
\frac{d\phi}{dr}\right)^2-V(\phi)\right], \label{eq:fe0}\\
\frac{1}{r}\frac{d\delta}{dr}&=&8\pi G\left[-\left(
\frac{d\phi}{dr}\right)^2\right], \\
\frac{1}{re^{-\delta}}\frac{d}{dr}
\left[re^{-\delta}\Delta\frac{d\phi}{dr}\right]&=&
\frac{\partial V}{\partial\phi}. 
\label{eq:fe00}
\end{eqnarray}

Now we will introduce dimensionless variables. We choose
\begin{equation}
\tilde{\phi}=\sqrt{8\pi G}\phi,~~~\tilde{r}=r/r_H,~~~
\tilde{\Delta}=\frac{\Delta}{\mu^2r_H^2},
\end{equation}
where $r_H$ is the radius of the horizon, i.e., $\Delta(r_H)=0$.
Using these variables, the field 
equations~(\ref{eq:fe0}-\ref{eq:fe00}) can be rewritten,
when $\phi<\phi_B$, as
\begin{eqnarray}
\frac{1}{\tilde{r}}\tilde{\Delta}'&=&-\tilde{\Delta}\left(
\tilde{\phi}'\right)^2+\tilde{\phi}^2, \label{eq:fe1}\\
\frac{1}{\tilde{r}}\delta'&=&-\left(
\tilde{\phi}'\right)^2, \\
\frac{1}{\tilde{r}e^{-\delta}}
\left[\tilde{r}e^{-\delta}\tilde{\Delta}\tilde{\phi}'
\right]'&+&\tilde{\phi}=0,
\label{eq:fe10}
\end{eqnarray}
where a prime (${}'$) denotes the derivative $d/d\tilde{r}$.

In the region $r_H<r<r_B$, the value of $\phi$ varies with
the radial coordinate, while 
$\phi$ takes the constant value
$\phi=\phi_B$ outside the boundary $r=r_B$.
Then the field equation for $\tilde{r}>\tilde{r}_B=r_B/r_H$ is
\begin{eqnarray}
\frac{1}{\tilde{r}}\tilde{\Delta}'&=&\tilde{\phi}_B^2, \\
\delta'&=&0, \\
\tilde{\phi}&=&\tilde{\phi}_B,
\label{eq:fe0out}
\end{eqnarray}
where $\tilde{\phi}_B=\sqrt{8\pi G}\phi_B$.
At the boundary $\tilde{r}=\tilde{r}_B$,
the interior and exterior solutons must be smoothly continued:
thus $\tilde{\phi}'=0$ at $\tilde{r}=\tilde{r}_B$.

%%%%%%%%%%%%%%%%%%%%%%%%%%%%%%%%%%%%%%%%%%%%%%%%%%%%%%%%%%%%%%%%%%%%%%

To solve the differential equations~(\ref{eq:fe1}-\ref{eq:fe10}), 
one must find
a suitable set of boudary conditions for a black hole solution.
A natural choice of the starting point is $\tilde{r}=1$.
This point corresponds to the horizon of the black hole, i.e.,
$\tilde{\Delta}(1)=0$. The value of $\delta$ moves obviously within a 
finite range. In the exterior region $\tilde{r}>\tilde{r}_B$, $\delta$ 
takes a constant value. In the region, the spacetime is the vacuum 
solution of pure gravity with an effective cosmological constant 
$-(8\pi G)\mu^2\phi_B^2/2$.
A constant shift of $\delta$ can be absorbed by redefinition of $t$.
Thus we choose $\delta(1)=0$.

We can arbitrarily choose $\tilde{\phi}_H=\tilde{\phi}(1)<0$.
Then we find that the condition for $\tilde{\phi}'(1)$ is
\begin{equation}
\tilde{\phi}'(1)=-\frac{\tilde{\phi}_H}{\tilde{\Delta}'(1)}.
\end{equation}
As a consequence of Eq.~(\ref{eq:fe1}), this equation is reduced to
\begin{equation}
\tilde{\phi}'(1)=-\frac{1}{\tilde{\phi}_H}.
\end{equation}

%%%%%%%%%%%%%%%%%%%%%%%%%%%%%%%%%%%%%%%%%%%%%%%%%%%%%%%%%%%%%%%%%%%%%%

Under these conditions, we can solve the 
equations~(\ref{eq:fe1}-\ref{eq:fe10})
numerically. If we choose a value of $\tilde{\phi}_H$, $\tilde{r}_B$ 
is determined by a numerical calculation: 
it is the value of the radial coordinate which fulfills
$\tilde{\phi}'(\tilde{r}_B)=0$.
$\tilde{\phi}_B$ is given by the value $\tilde{\phi}(\tilde{r}_B)$.

\begin{figure}[h]
\centering
\epsfbox{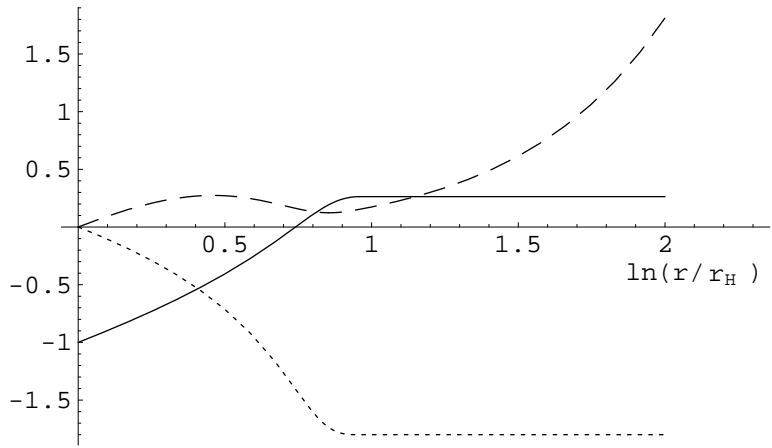}
\caption{The solutions of $\tilde{\phi}$ (the solid line), 
$\tilde{\Delta}$ (the broken line), and 
$\delta$ (the dotted line) for $\tilde{\phi}_H=-1$.}
\label{fig:2}
\end{figure}

The numerical solutions of $\tilde{\phi}$, $\tilde{\Delta}$, and 
$\delta$ for a specific value 
$\tilde{\phi}_H=-1$ are shown in FIG.~\ref{fig:2}.

%%%%%%%%%%%%%%%%%%%%%%%%%%%%%%%%%%%%%%%%%%%%%%%%%%%%%%%%%%%%%%%%%%%%%%

The properties of a black hole solution are extracted from the value 
of the variables at $\tilde{r}=\tilde{r}_B$.
In the exterior region, the variables are ones of the BTZ black hole 
solution except for the scaling of the time coordinate due to 
$\delta$.
The smooth connection at the boundary determines the mass of the black
hole $M$. We obtain:
\begin{equation}
\frac{8GM}{\mu^2 r_H^2}=\frac{1}{2}\tilde{\phi}_B^2\tilde{r}_B^2-
\tilde{\Delta}_B,
\end{equation}
where $\tilde{\Delta}_B=\tilde{\Delta}(\tilde{r}_B)$.
The effective cosmological constant is given by
\begin{equation}
\frac{\lambda}{\mu^2}=\frac{1}{2}\tilde{\phi}_B^2.
\end{equation}

%%%%%%%%%%%%%%%%%%%%%%%%%%%%%%%%%%%%%%%%%%%%%%%%%%%%%%%%%%%%%%%%%%%%%%

The black hole solution has the three characteristic length scale;
$r_H$, $r_B$, and the horizon radius of the BTZ black hole which has 
the same mass and cosmological constant, i.e.,
\begin{equation}
r_{H0}=\sqrt{\frac{8GM}{\lambda}}.
\end{equation}

\begin{figure}[h]
\centering
\epsfbox{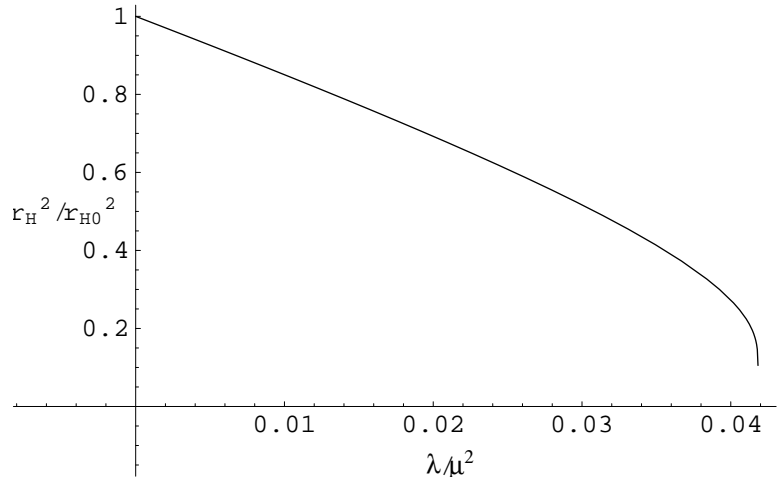}
\caption{$r_H^2/r_{H0}^2$ as a function of $\lambda/\mu^2$.}
\label{fig:3}
\end{figure}

\begin{figure}[h]
\centering
\epsfbox{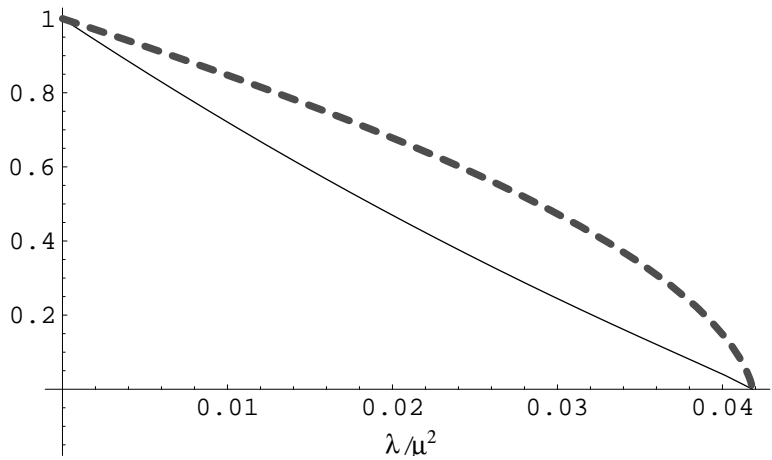}
\caption{$r_H^2/r_{B}^2$ (the solid line) and 
$r_{H0}^2/r_{B}^2$ (the gray broken line) as functions
 of $\lambda/\mu^2$.}
\label{fig:4}
\end{figure}

The ratios of $r_H^2/r_{H0}^2$ is plotted as a function of 
$\lambda/\mu^2$ in FIG.~\ref{fig:3}.
In FIG.~\ref{fig:4}, $r_H^2/r_{B}^2$ and $r_{H0}^2/r_{B}^2$ are shown
as functions of $\lambda/\mu^2$.
For a large value of $\lambda/\mu^2$, the present type of the black 
hole cannnot exist. The critical value is approximately given by
$\lambda/\mu^2\approx 0.042$.

On the other hand, in the limit of small $\lambda/\mu^2$, all the 
ratios converge to unity. Then the effect of ``scalar hair'' tends to
be small and the black hole approaches a usual BTZ black hole
in the limit $\lambda/\mu^2\rightarrow 0$.

%%%%%%%%%%%%%%%%%%%%%%%%%%%%%%%%%%%%%%%%%%%%%%%%%%%%%%%%%%%%%%%%%%%%%%

The Hawking temperature of the black hole
is proportional to the strength of the 
surface gravity at the horizon.
Practically speaking, the Hawking temperature $T$ is derived from 
the condition that the Euclideanized metric has no conical singularity
at the horizon when the Euclidean time has the period $1/T$. In the 
present case, $T$ is given as
\begin{equation}
\frac{T}{\mu^2 r_H}=\frac{1}{4\pi}e^{\delta_B}\tilde{\phi}_H^2.
\end{equation}
Here the dependence on $\delta_B=\delta(\tilde{r}_B)$ comes from
the redefinition of $t$, which converts the exterior solution into
the BTZ solution explicitly. 

\begin{figure}[h]
\centering
\epsfbox{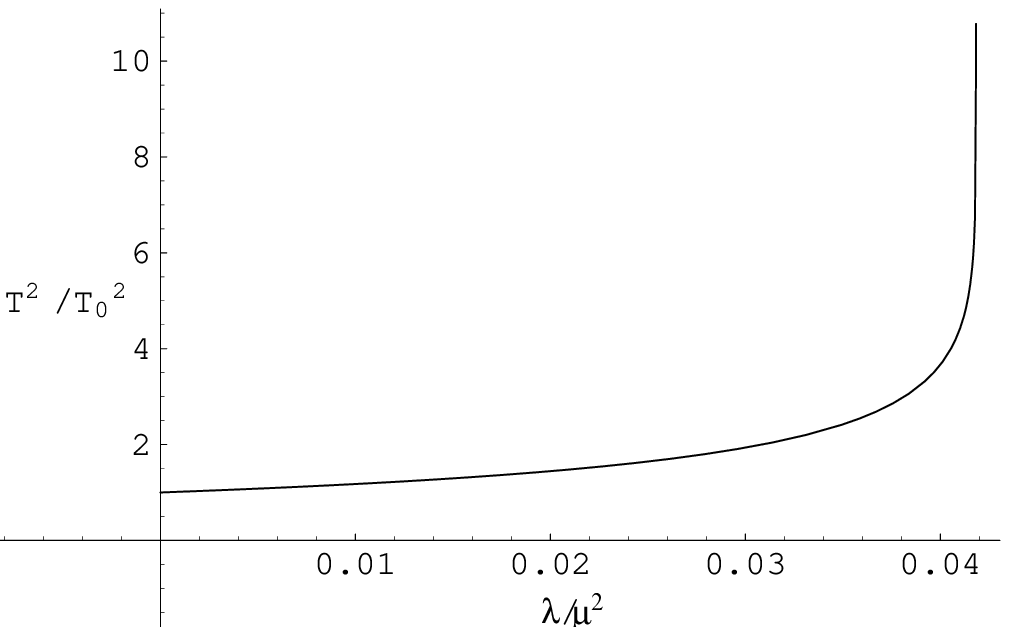}
\caption{$T^2/T_{0}^2$ as a function of $\lambda/\mu^2$.}
\label{fig:5}
\end{figure}

In FIG.~\ref{fig:5}, we show the ratio $T^2/T_0^2$,
where $T_0$ is the Hawking temperature of the BTZ black hole of the 
same mass and cosmological constant~\cite{BTZ}:
\begin{equation}
T_0=\frac{\sqrt{8GM\lambda}}{2\pi}.
\end{equation}

The ratio $T^2/T_0^2$ is always larger than unity 
for a finite value of $\lambda/\mu^2$ less than the critical value.
The ratio grows up if the value of $\lambda/\mu^2$ approaches to
the critical value, $\lambda/\mu^2\approx 0.042$. In the limit of 
small $\lambda/\mu^2$, the ratio becomes unity.

%%%%%%%%%%%%%%%%%%%%%%%%%%%%%%%%%%%%%%%%%%%%%%%%%%%%%%%%%%%%%%%%%%%%%%
%%%%%%%%%%%%%%%%%%%%%%%%%%%%%%%%%%%%%%%%%%%%%%%%%%%%%%%%%%%%%%%%%%%%%%
\section{Rotating black holes with scalar hair}
\label{sec:3}
%%%%%%%%%%%%%%%%%%%%%%%%%%%%%%%%%%%%%%%%%%%%%%%%%%%%%%%%%%%%%%%%%%%%%%
%%%%%%%%%%%%%%%%%%%%%%%%%%%%%%%%%%%%%%%%%%%%%%%%%%%%%%%%%%%%%%%%%%%%%%

For the rotating, circularly-symmetric solutions,
the metric can be taken as 
\begin{equation}
ds^{2}= - e^{-2\delta(r)}\Delta(r)dt^{2} + \frac{1}{\Delta(r)} dr^{2}+
r^2\left(d\theta-\Omega(r)dt\right)^2,
\label{eq:met1}
\end{equation}

We also assume $\phi$ is the function of the radial coodinate.
Note that under the circularly-symmetric ansatz, a real scalar
cannot depend on the angular variable.

One can obtain the field equations by taking variations
of the action~(\ref{eq:action}) under the circularly-symmetric ansatz.
Since the $(0,2)$ component of Einstein equation does not include
$\phi$ dependence in our case, the equation can be integrated and 
leads to
\begin{equation}
e^{2\delta}\left(\frac{d\Omega}{dr}\right)^2=
\frac{(8GJ)^2}{r^6},
\end{equation}
where $J$ is a constant. $J$ turns out to be the value of 
the angular momentum of the black hole\cite{BTZ}.

Other field equations can be rewritten by using the dimensionless
variables as in the previous section. In this time we take
\begin{equation}
\tilde{\phi}=\sqrt{8\pi G}\phi,~~~\tilde{r}=r/r_H,~~~
\tilde{\Delta}=\frac{\Delta}{\mu^2r_H^2},~~~and~~~
\tilde{J}=\frac{8GJ}{\mu r_H^2}.
\end{equation}

Using these variables, the field equations for $\phi<\phi_B$ 
can be read as
\begin{eqnarray}
\frac{1}{\tilde{r}}\tilde{\Delta}'&=&-\tilde{\Delta}\left(
\tilde{\phi}'\right)^2+\tilde{\phi}^2-
\frac{\tilde{J}^2}{2\tilde{r}^4}, 
\label{eq:fe2}\\
\frac{1}{\tilde{r}}\delta'&=&-\left(
\tilde{\phi}'\right)^2, \\
\frac{1}{\tilde{r}e^{-\delta}}
\left[\tilde{r}e^{-\delta}\tilde{\Delta}\tilde{\phi}'
\right]'&+&\tilde{\phi}=0. 
\label{eq:fe20}
\end{eqnarray}

%%%%%%%%%%%%%%%%%%%%%%%%%%%%%%%%%%%%%%%%%%%%%%%%%%%%%%%%%%%%%%%%%%%%%%

As in the nonrotating case, $\phi$ takes the constant value
$\phi=\phi_B$ in the exterior region of the boundary $r=r_B$. Then the
field equation for $\tilde{r}>\tilde{r}_B=r_B/r_H$ is
\begin{eqnarray}
\frac{1}{\tilde{r}}\tilde{\Delta}'&=&\tilde{\phi}_B^2-
\frac{\tilde{J}^2}{2\tilde{r}^4}, \\
\delta'&=&0, \\
\tilde{\phi}&=&\tilde{\phi}_B,
\label{eq:fe1out}
\end{eqnarray}
where $\tilde{\phi}_B=\sqrt{8\pi G}\phi_B$.
The smooth connection of
the interior and exterior solutons must be required.

A suitable set of boudary conditions for solving the differential 
equations~(\ref{eq:fe2}-\ref{eq:fe20}) can be found as in the previous
case. At $\tilde{r}=1$, we set $\tilde{\Delta}(1)=0$, $\delta(1)=0$, 
and
\begin{eqnarray}
\tilde{\phi}'(1)&=&-\frac{\tilde{\phi}_H}{\tilde{\Delta}'(1)} \\
&=&-\frac{\tilde{\phi}_H}{\tilde{\phi}_H^2-\frac{\tilde{J}^2}{2}}.
\end{eqnarray}

%%%%%%%%%%%%%%%%%%%%%%%%%%%%%%%%%%%%%%%%%%%%%%%%%%%%%%%%%%%%%%%%%%%%%%

The properties of a rotating black hole solution can be extracted from
the numerical value of the variables at $\tilde{r}=\tilde{r}_B$ as 
previously.
In the exterior region, the solution is the rotating BTZ black hole 
solution~\cite{BTZ}. By examining the connection condition, we find:
\begin{equation}
\frac{8GM}{\mu^2 r_H^2}=\frac{1}{2}\tilde{\phi}_B^2\tilde{r}_B^2+
\frac{\tilde{J}^2}{4\tilde{r}_B^2}-\tilde{\Delta}_B,
\end{equation}
and the effective cosmological constant is again given by
\begin{equation}
\frac{\lambda}{\mu^2}=\frac{1}{2}\tilde{\phi}_B^2.
\end{equation}

The horizon radius of the rotating BTZ black hole with 
the same mass, angular momentum, and cosmological constant is
given by
\begin{equation}
r_{H0}=\sqrt{\frac{1}{2}\left(
\frac{\bar{M}}{\lambda}+
\sqrt{\frac{\bar{M}^2}{\lambda^2}-
\frac{\bar{J}^2}{\lambda}}\right)},
\end{equation}
where
\begin{equation}
\bar{M}=8GM~~~and~~~\bar{J}=8GJ.
\end{equation}

\begin{figure}[h]
\centering
\epsfbox{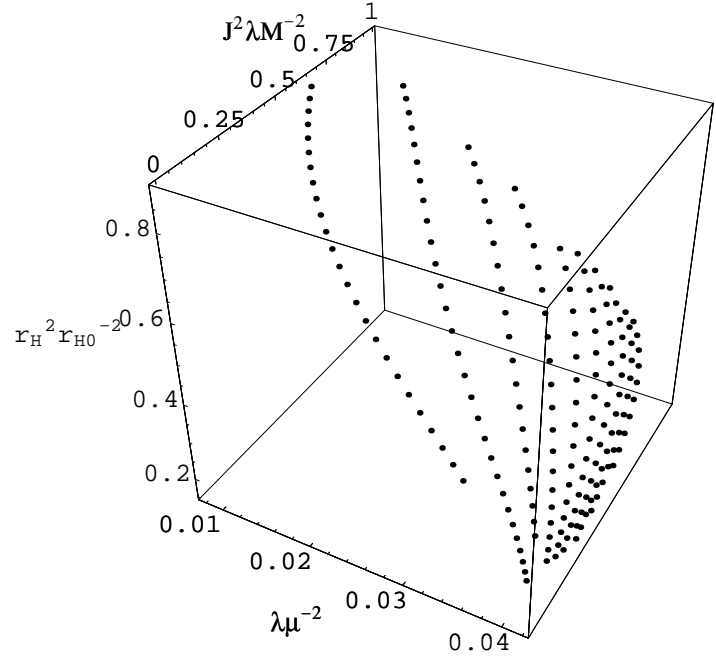}
\caption{$r_H^2/r_{H0}^2$ as a function of $\lambda/\mu^2$
and $J^2\lambda/M^2$
for rotating black holes.}
\label{fig:6}
\end{figure}

\begin{figure}[h]
\centering
\epsfbox{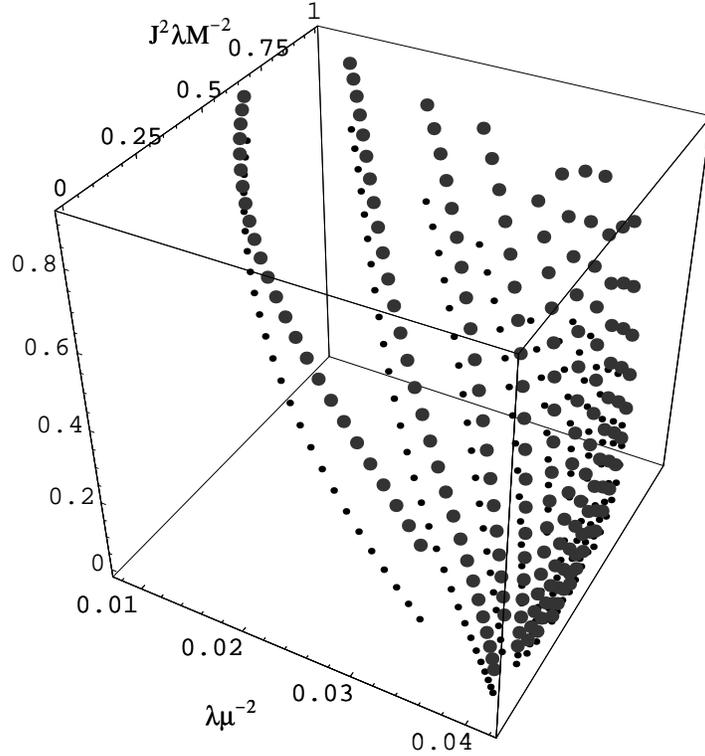}
\caption{$r_H^2/r_{B}^2$ and $r_{H0}^2/r_{B}^2$ as functions
 of $\lambda/\mu^2$ 
and $J^2\lambda/M^2$ for rotating black holes. 
The small dots represent $r_H^2/r_{B}^2$,  
while the large gray dots represent $r_{H0}^2/r_{B}^2$.}
\label{fig:7}
\end{figure}

The ratios $r_H^2/r_{H0}^2$ for rotating black holes are plotted
in FIG.~\ref{fig:6}. In FIG.~\ref{fig:7}, the values of
$r_H^2/r_{B}^2$ and $r_{H0}^2/r_{B}^2$ for rotating black holes
are plotted.

In these figures, the $x$-axis indicates $\lambda/\mu^2$
while $y$-axis $J^2\lambda/M^2$.
In these figures, the sequences of the points correspond to
$\frac{8GJ}{\mu r_H}=0.1,0.2,0.3,0.4,0.5,0.6,0.7,0.8,0.9,$ and $1$.

It is worth noting that the scattered points, which correspond to 
different solutions, lie within the range $0<J^2\lambda/M^2<1$.
The same restriction hold for the rotating BTZ black hole solution
(with the horizon)~\cite{BTZ}.

\begin{figure}[h]
\centering
\epsfbox{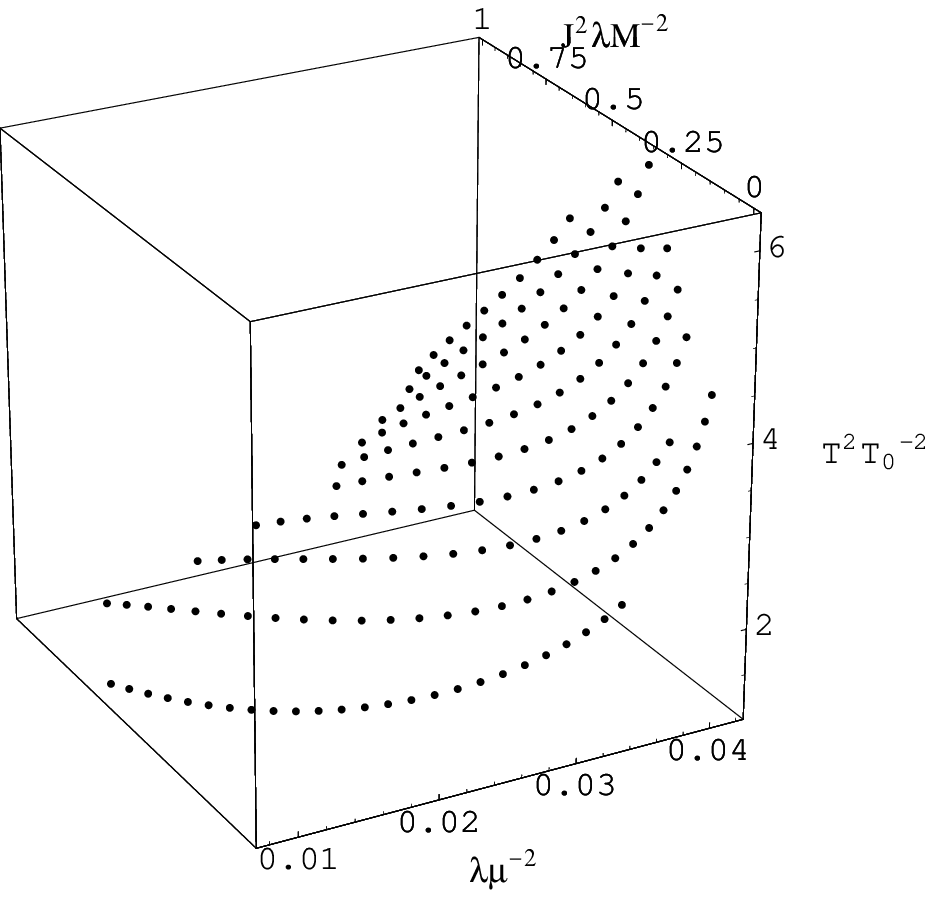}
\caption{$T^2/T_{0}^2$ as a function of $\lambda/\mu^2$
and $J^2\lambda/M^2$
for rotating black holes.}
\label{fig:8}
\end{figure}

In the rotating case, the Hawking temperature $T$ is given by
\begin{equation}
\frac{T}{\mu^2 r_H}=\frac{1}{4\pi}e^{\delta_B}\left(
\tilde{\phi}_H^2-\frac{\tilde{J}^2}{2}\right).
\end{equation}
In FIG.~\ref{fig:8}, we show the ratio $T^2/T_0^2$,
where $T_0$ is the Hawking temperature of the BTZ black hole of the 
same mass, angular momentum and cosmological constant~\cite{BTZ}:
\begin{equation}
T_0=\frac{\lambda}{2\pi r_{H0}}
\sqrt{\frac{\bar{M}^2}{\lambda^2}-
\frac{\bar{J}^2}{\lambda}}.
\end{equation}
Again the $x$-axis indicates $\lambda/\mu^2$
while $y$-axis $J^2\lambda/M^2$ in FIG.~\ref{fig:8}.

\begin{figure}[h]
\centering
\epsfbox{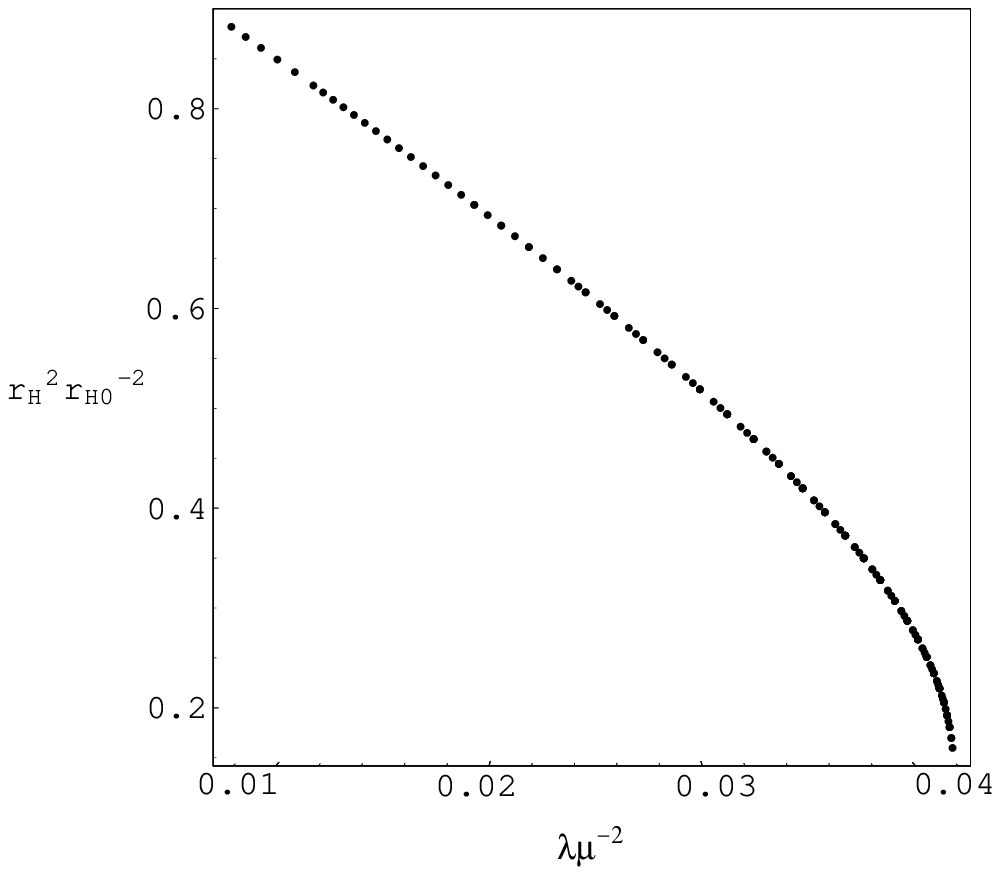}
\caption{$r_H^2/r_{H0}^2$ as a function of $\lambda/\mu^2$
for rotating black holes. This is the same as FIG.~6, but
projected onto a plane.}
\label{fig:9}
\end{figure}

\begin{figure}[h]
\centering
\epsfbox{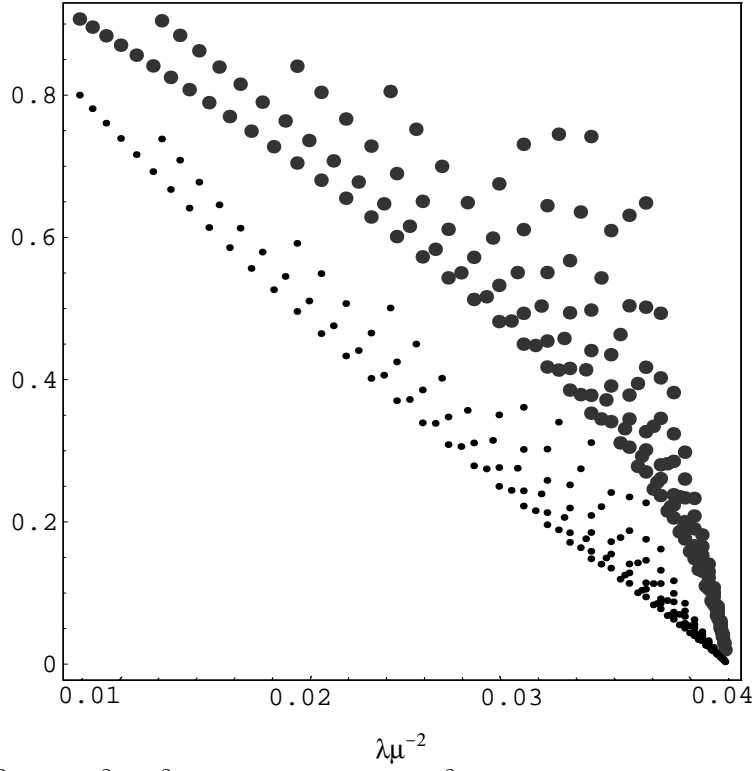}
\caption{$r_H^2/r_{B}^2$ and $r_{H0}^2/r_{B}^2$ as functions
of $\lambda/\mu^2$ for rotating black holes. The small dots represent
$r_H^2/r_{B}^2$,  
while the large gray dots represent $r_{H0}^2/r_{B}^2$. This
is the same as FIG.~7, but projected onto a plane.}
\label{fig:10}
\end{figure}

\begin{figure}[h]
\centering
\epsfbox{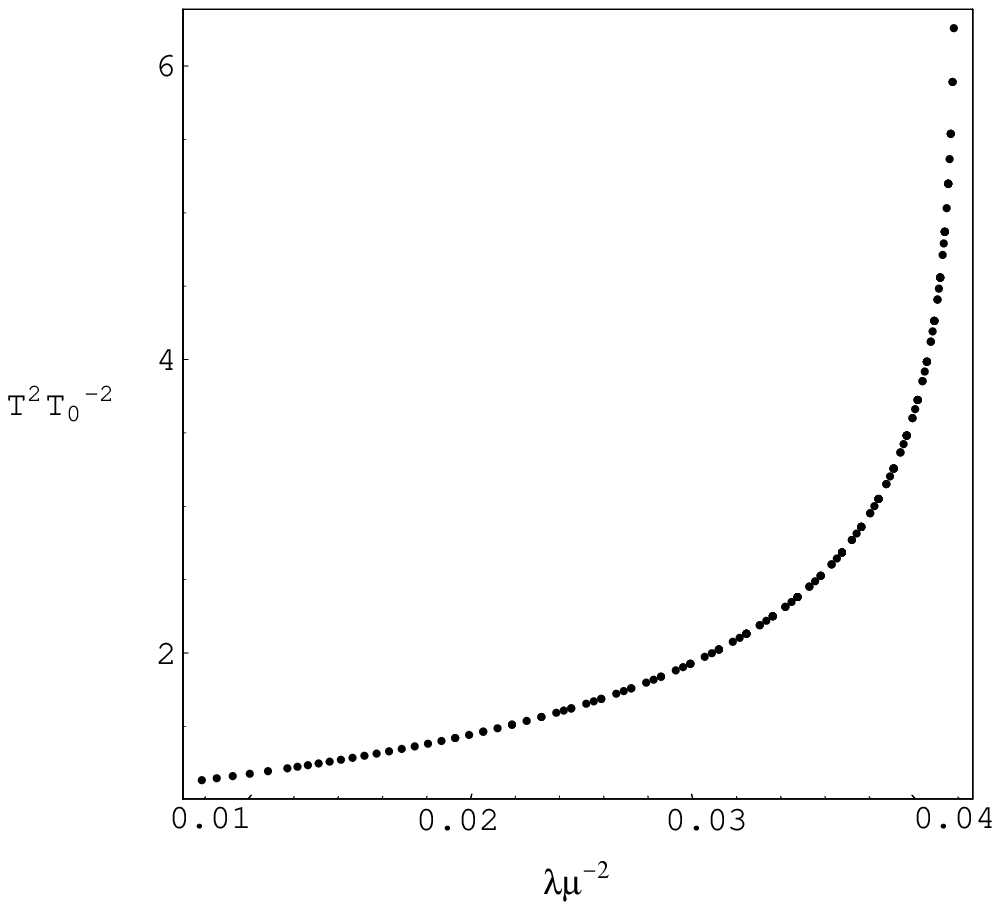}
\caption{$T^2/T_{0}^2$ as a function of $\lambda/\mu^2$ 
for rotating black holes. This is the same as FIG.~8, but
projected onto a plane.}
\label{fig:11}
\end{figure}

What does the result mean?
Now FIG.~\ref{fig:9} shows the scattered points in FIG.~\ref{fig:6} 
projected onto the $\lambda/\mu^2$-$r_H^2/r_{H0}^2$ plane.
The points are laid on a curve, which is the same curve
in the nonrotating case.
Similarly, in FIG.~\ref{fig:10} the projected points of 
FIG.~\ref{fig:7} are shown. The ratios $r_H^2/r_{B}^2$ and 
$r_{H0}^2/r_{B}^2$ depend on the ratio of the angular momentum and 
mass of the black hole. In FIG.~\ref{fig:11} the projected points of
FIG.~\ref{fig:8} are shown. The ratio $T^2/T_0^2$ does not depend on
the ratio of the angular momentum and mass of the black hole.

In summary, $r_H^2/r_{H0}^2$ and $T^2/T_0^2$
are independent of the value of angular momentum.
Therefore the critical value of $\lambda/\mu^2$ is the same as 
the one of the nonrotating case, $\lambda/\mu^2\approx 0.042$.
On the other hand, $r_H^2/r_{B}^2$ and $r_{H0}^2/r_{B}^2$ depend on
the ratio of the angular momentum and mass of the black hole.
For larger angular momenta, the ratios become larger.

%%%%%%%%%%%%%%%%%%%%%%%%%%%%%%%%%%%%%%%%%%%%%%%%%%%%%%%%%%%%%%%%%%%%%%
%%%%%%%%%%%%%%%%%%%%%%%%%%%%%%%%%%%%%%%%%%%%%%%%%%%%%%%%%%%%%%%%%%%%%%
\section{Conclusion}
\label{sec:4}
%%%%%%%%%%%%%%%%%%%%%%%%%%%%%%%%%%%%%%%%%%%%%%%%%%%%%%%%%%%%%%%%%%%%%%
%%%%%%%%%%%%%%%%%%%%%%%%%%%%%%%%%%%%%%%%%%%%%%%%%%%%%%%%%%%%%%%%%%%%%%

We have constructed circularly-symmetric black hole solutions with
scalar hair in $(2+1)$ dimensions.
In our model, ratios of physical quantities have 
dependence on the scaled cosmological constant $\lambda/\mu^2$,
because of the simplicity of the scalar potential.
This fact tells us that the complexity in the behavior of the physical
quantities in the case of the Yang-Mills black holes in $(3+1)$ 
dimensions~\cite{Biz,Tor} is due to the non-linear nature of the 
self-interaction.

Although our model is very simple, we found the critical
behavior with respect to  $\lambda/\mu^2$.
We have to study more closely this behavior and
the structure of the spacetime when $\lambda/\mu^2$
approaches the critical value $\approx 0.042$.

For rotating black holes, we found some ratios of the 
physical quantities are independent of the angular momentum.
The extreme condition seems the same as the one of the vacuum case,
$J^2\lambda/M^2=1$. Studying the structure of the spacetime in the 
extreme case is of much interest.

Analyzing the stability of our solution is, unfortunately,
somewhat difficult because of the singular point in the potential.
The general cases with ``smooth'' potentials must be investigated 
and the 
relation to the model of the unified field theory has to be clarified.

%%%%%%%%%%%%%%%%%%%%%%%%%%%%%%%%%%%%%%%%%%%%%%%%%%%%%%%%%%%%%%%%%%%%%%
%%%%%%%%%%%%%%%%%%%%%%%%%%%%%%%%%%%%%%%%%%%%%%%%%%%%%%%%%%%%%%%%%%%%%%

%%%%%%%%%%%%%%%%%%%%%%%%%%%%%%%%%%%%%%%%%%%%%%%%%%%%%%%%%%%%%%%%%%%%%%
%%%References
%%%%%%%%%%%%%%%%%%%%%%%%%%%%%%%%%%%%%%%%%%%%%%%%%%%%%%%%%%%%%%%%%%%%%%

\newpage

%%%%%%%%%%%%%%%%%%%%%%%%%%%%%%%%%%%%%%%%%%%%%%%%%%%%%%%%%%%%%%%%%%%%%%
%%%FIG. 1
%%%%%%%%%%%%%%%%%%%%%%%%%%%%%%%%%%%%%%%%%%%%%%%%%%%%%%%%%%%%%%%%%%%%%%

%%%%%%%%%%%%%%%%%%%%%%%%%%%%%%%%%%%%%%%%%%%%%%%%%%%%%%%%%%%%%%%%%%%%%%
%%%FIG. 2
%%%%%%%%%%%%%%%%%%%%%%%%%%%%%%%%%%%%%%%%%%%%%%%%%%%%%%%%%%%%%%%%%%%%%%

%%%%%%%%%%%%%%%%%%%%%%%%%%%%%%%%%%%%%%%%%%%%%%%%%%%%%%%%%%%%%%%%%%%%%%
%%%FIG. 3
%%%%%%%%%%%%%%%%%%%%%%%%%%%%%%%%%%%%%%%%%%%%%%%%%%%%%%%%%%%%%%%%%%%%%%

%%%%%%%%%%%%%%%%%%%%%%%%%%%%%%%%%%%%%%%%%%%%%%%%%%%%%%%%%%%%%%%%%%%%%%
%%%FIG. 4
%%%%%%%%%%%%%%%%%%%%%%%%%%%%%%%%%%%%%%%%%%%%%%%%%%%%%%%%%%%%%%%%%%%%%%

%%%%%%%%%%%%%%%%%%%%%%%%%%%%%%%%%%%%%%%%%%%%%%%%%%%%%%%%%%%%%%%%%%%%%%
%%%FIG. 5
%%%%%%%%%%%%%%%%%%%%%%%%%%%%%%%%%%%%%%%%%%%%%%%%%%%%%%%%%%%%%%%%%%%%%%

%%%%%%%%%%%%%%%%%%%%%%%%%%%%%%%%%%%%%%%%%%%%%%%%%%%%%%%%%%%%%%%%%%%%%%
%%%FIG. 6
%%%%%%%%%%%%%%%%%%%%%%%%%%%%%%%%%%%%%%%%%%%%%%%%%%%%%%%%%%%%%%%%%%%%%%

%%%%%%%%%%%%%%%%%%%%%%%%%%%%%%%%%%%%%%%%%%%%%%%%%%%%%%%%%%%%%%%%%%%%%%
%%%FIG. 7
%%%%%%%%%%%%%%%%%%%%%%%%%%%%%%%%%%%%%%%%%%%%%%%%%%%%%%%%%%%%%%%%%%%%%%

%%%%%%%%%%%%%%%%%%%%%%%%%%%%%%%%%%%%%%%%%%%%%%%%%%%%%%%%%%%%%%%%%%%%%%
%%%FIG. 8
%%%%%%%%%%%%%%%%%%%%%%%%%%%%%%%%%%%%%%%%%%%%%%%%%%%%%%%%%%%%%%%%%%%%%%

%%%%%%%%%%%%%%%%%%%%%%%%%%%%%%%%%%%%%%%%%%%%%%%%%%%%%%%%%%%%%%%%%%%%%%
%%%FIG. 9
%%%%%%%%%%%%%%%%%%%%%%%%%%%%%%%%%%%%%%%%%%%%%%%%%%%%%%%%%%%%%%%%%%%%%%

%%%%%%%%%%%%%%%%%%%%%%%%%%%%%%%%%%%%%%%%%%%%%%%%%%%%%%%%%%%%%%%%%%%%%%
%%%FIG. 10
%%%%%%%%%%%%%%%%%%%%%%%%%%%%%%%%%%%%%%%%%%%%%%%%%%%%%%%%%%%%%%%%%%%%%%

%%%%%%%%%%%%%%%%%%%%%%%%%%%%%%%%%%%%%%%%%%%%%%%%%%%%%%%%%%%%%%%%%%%%%%
%%%FIG. 11
%%%%%%%%%%%%%%%%%%%%%%%%%%%%%%%%%%%%%%%%%%%%%%%%%%%%%%%%%%%%%%%%%%%%%%

%%%%%%%%%%%%%%%%%%%%%%%%%%%%%%%%%%%%%%%%%%%%%%%%%%%%%%%%%%%%%%%%%%%%%%
\end{document}